# Locking and Tracking Magnetic Resonance Spectra of NV- Center for Real-time Magnetometry using a Differential Photon-Rate Meter

Kapildeb Ambal[1,2], Robert D. McMichael[1]

[1]Center for Nanoscale Science and Technology, National Institute of Standards and Technology

[2]Institute for Research in Electronics and Applied Physics, University of Maryland, College Park, MD 20742

**Abstract:**

We describe a real-time data processing and frequency control method to track peaks in optically detected magnetic resonance of nitrogen-vacancy centers in diamond. This procedure allows us to measure magnetic field continuously with sensitivity ≈ 6 µT/Hz$^{1/2}$ and to track resonances at sweep rates exceeding 110 µT/s. We use a custom-built differential photon rate meter and active feedback control to "lock" the microwave excitation frequency of the magnetic resonance. Our scheme covers a broad magnetic field range, limited by the frequency range of the microwave generator. This measurement procedure automates the processing of voltage pulse outputs from the photon counters, and it provides sensitivity comparable to traditional photon counting methods.

**Introduction:**

Real-time and precise measurement of magnetic field is important for many scientific and technological applications ranging from physics to biology [1,2], and a variety of sensors have developed to meet different needs. Superconducting quantum interference devices (SQUIDs) can measure extremely weak magnetic fields [3][4], but they have to be cooled to cryogenic temperatures. Hall probes conveniently measure relatively large magnetic fields but suffer from offsets that require occasional calibration. However, magnetic resonance based sensors [5][6][7][8], such as nuclear magnetic resonance (NMR) magnetometers, nitrogen vacancy (NV) centers in diamond [9][10][11], and organic polymer based magnetometers [5][6] allow drift-free and offset-free determination of magnetic fields over a range of temperatures.

A feature of magnetic resonance based sensors is that their operation depends strongly on fundamental physical constants, and only weakly on extrinsic quantities like temperature and materials properties. Generally, magnetic resonance based magnetometers exploit Planck's fundamental relationship between the frequency ($\nu$) of electromagnetic radiation and Zeeman energy ($h\nu = h\gamma B_0$) of the paramagnetic centers to determine the magnetic field [12][13] ($\gamma$ is the gyromagnetic ratio, $B_0$ is the applied magnetic field and $h$ is the Planck constant). Since the gyromagnetic ratio is independent of both temperature and magnetic fields, magnetic resonance-based sensors are better alternatives for absolute field calibrations. The nitrogen vacancy center in diamond is a magnetic resonance based sensor that is particularly useful for precise magnetic field mapping with nanoscale spatial resolution. The NV- centers' small size (2 atoms) and excellent field sensitivity (≈1 nT/Hz$^{1/2}$) at room temperature have made them increasingly popular for nanoscale magnetic field mapping over mesoscopic and macroscopic areas [14][15].

NV- center-based magnetic field sensing schemes are usually either continuous-wave optically detected magnetic resonance (cw-ODMR) or pulsed ODMR. Although pulsed ODMR measurements provide better sensitivity[16][17][18], we concentrate on cw-ODMR here. The cw-ODMR method is less technically demanding, making it more convenient for sensing static and low-frequency fields and for measuring spatial field variations.

The noise floor for field measurements using cw-ODMR is largely determined by noise in the fluorescence signal and by the signal sensitivity $dS/dB$, which in turn depends on the fluorescence contrast and the resonance linewidth. Thus, the challenge in maximizing signal sensitivity lies in generating the narrowest spectral linewidth during cw-excitation while ensuring that the fluorescence contrast is as high as possible. The linewidth depends on the inherent collective-spin dephasing rates, as well as the optical and microwave (MW) power-related broadening[19]. The intrinsic properties of NV- centers can be improved by using isotopically pure carbon ($^{12}$C) and the spectral linewidth can be tuned by selecting isotopically pure $^{15}$N as compared to $^{14}$N, but these discussions are beyond the scope of this paper[20][21][22][23][18]. In the best case, the dominant noise source is the shot noise of single photon detection, which is primarily limited by the number of NV- centers used and the amount of light that is generated and collected. In addition to the intrinsic shot noise, extrinsic noise factors include environmental drift, technical noise of the detection apparatus and fluctuations in both laser power and microwave power.

Phase-sensitive detection or "lock-in" techniques are frequently used to avoid many of these extrinsic noise sources. The lock-in amplifier can make best use of its dynamic range when the input is a smoothly varying AC voltage. In the case of a large ensembles of NV- centers, these signals are available, as the emitted photoluminescence power is within the detection range of regular photo-detectors. Very large ensembles of NV- centers (>$10^6$ NV- centers) have produced sub-nanotesla sensitivity in work employing regular photo-detectors and lock-in detection [9][24][25][26][27][23]. The large number of NV- centers improves detection signals, but the density of defects must be kept low enough to avoid line broadening, and the resulting reduction in field sensitivity. As a result, a large ensemble NV- magnetometer is a good alternative and when spatial resolution is not so important [18,23].

On the other hand, in cases where a single NV- center or a few NV- centers are used, for example to achieve better spatial resolution, the collected photon output power is ≈ 10 fW, which is below the detection limit of regular photo-detectors. Instead, the photons are often detected using avalanche photo diode detectors (APD). The APD generates a narrow (≈ 20 ns), discrete voltage pulse for each detected photon and typical photon count rates are in the range of $10^5$ s$^{-1}$ to $10^6$ s$^{-1}$. Unfortunately, this train of pulses is not compatible with typical lock-in amplifiers. Instead, the conventional approach is to count pulses and communicate the results to a computer for post-processing [28] [11] [29].

In addition to avoiding extrinsic noise, lock-in detection also facilitates field tracking by producing a continuous measurement signal. The goal of the field tracking is to find and follow the center of a resonance peak. (The slower, but more precise method would be to measure and fit a spectrum.) For tracking applications, frequency modulation with phase-sensitive detection provides a DC error signal that can be used to "lock" the resonance peak with feedback control of the excitation frequency. This scheme allows rapid field measurements in cases where the field varies by small amounts between measurements. Such active feedback control for real-time magnetometry has been demonstrated[9][30][24] using large ensembles of NV- centers, which emit relatively high photoluminescence power measurable by regular photo-diodes. However, there are few reports of active feedback control detection schemes and real-time magnetometry using single NV- centers or small ensembles of NV- centers, which emit relatively low photoluminescence power requiring single photon detection[28] [11] .

When single photon-detectors (APD) are needed, it becomes a greater challenge to convert the detector's output pulses into a smooth continuous signal to be used for phase sensitive detection and peak tracking. Previous work [11] demonstrating peak tracking using single-photon-detector (APD) input used photon pulse counting with data transmission to a computer and demodulation and feedback control via computer algorithms.

Here, we address active feedback control and field tracking using frequency modulated differential (lock-in-type) signal processing with photo detection rates in the range from 2 ms$^{-1}$ to 500 ms$^{-1}$ (≈50 fW). In the Experimental Procedure section below, we describe our test setup, including a custom differential rate meter with phase sensitive detection that generates a normalized difference signal from APD pulse inputs. Following that, the Results and Discussion section presents an evaluation of the signal-to-noise ratio, sensitivity (6 µT/Hz$^{1/2}$), and field tracking for field sweep rates exceeding 100 µT/s.

## Experimental Procedure:

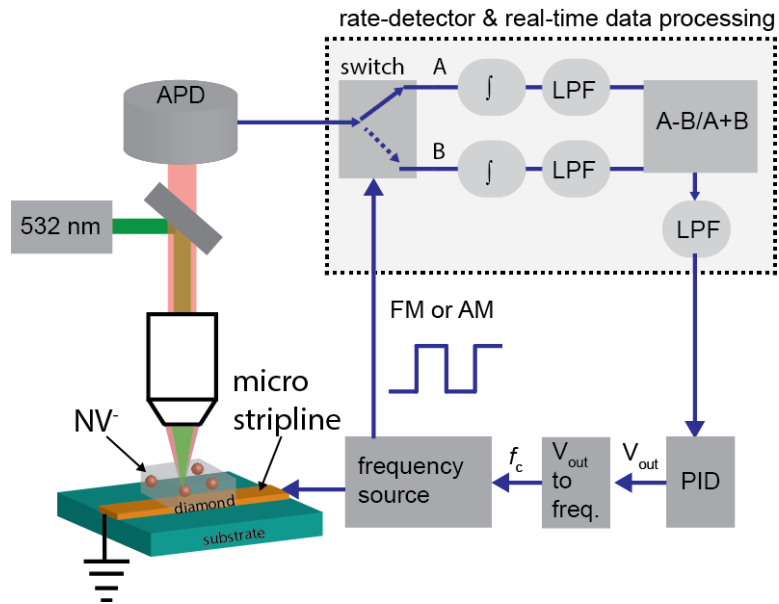

*Fig.1 Magnetic resonance tracking system using NV⁻ centers as magnetic field sensors. NV⁻ centers are excited with green laser light and emitted photoluminescence is collected using an avalanche photo diode (APD). The APD signal is sent to the differential rate meter, which contains two gated high-speed voltage pulse integrators, a low pass filter (LPF) and an analog operation for a normalized differential output. The PID controller monitors any changes in the output of the rate meter and generates a compensating frequency-control output that tracks the ODMR resonance peak.*

The experimental setup consists of three parts; 1) confocal microscope for optically detected magnetic resonance (ODMR) measurement using NV⁻ center, 2) differential rate meter, and 3) frequency locking using a proportional-integral-derivative (PID) controller for frequency locking.

The confocal microscope for ODMR measurement consists of 1) a green laser for excitation, 2) a custom-built electromagnet for the applied magnetic field, and 3) a microwave antenna to drive spin resonance. The NV⁻ centers are continuously excited with green laser light (532 nm) with nominal illumination ≈100 µW, which is low enough to avoid optical power related broadening. Emitted photoluminescence (600 nm to 800 nm) is collected and detected by an avalanche photo diode (APD). The APD outputs a 20 ns TTL pulse per detected photon. For ODMR measurement, the photon detection rate is then recorded using the rate meter as a function of microwave frequency.

To demodulate a modulated photoluminescence signal, the rate meter switches the input pulse train from the APD between the inputs of two parallel channels synchronously with the modulation. In each channel, pulses are made uniform in amplitude and duration, and then integrated in an frequency-to-voltage circuit based on a voltage controlled oscillator with negative feedback regulation. These integrators function at pulse rates from 2 ms$^{-1}$ to 500 ms$^{-1}$. Finally, the normalized difference is obtained from the individual channel voltages using an analog multiplier/divider chip.

The rate meter is used with either amplitude modulation or frequency modulation. For amplitude modulated (AM) cw-ODMR measurements, the microwave power alternates between ON and OFF states. In the OFF state, the NV$^-$ centers are continuously pumped into the bright $|0\rangle$ spin state, while in the ON state, photoluminescence is reduced as spins are driven into $|-1\rangle$ and $|+1\rangle$ states. The normalized differential output (A-B)/(A+B) is recorded as a function of the microwave frequency. Results are shown in Figure 2.

For frequency modulated (FM) cw-ODMR measurements, which we use for peak tracking, the microwave frequency alternates between two frequencies, $f_1$ and $f_2$, around a center frequency, $f_c$. Channel-A opens while the frequency is $f_1$ and channel-B opens while the frequency is $f_2$ respectively. Similarly to the amplitude-modulated case, we record the frequency-modulated, normalized-difference, cw-ODMR spectrum as a function of microwave frequencies ($f_c$) as shown in figure-3a.

The derivative-like line shape of the frequency-modulated signal provides an error signal for locking the resonance peak. Near resonance, the signal is positive if the driving frequency is above resonance, and negative if the driving frequency is below resonance. We use active feedback to generate a frequency-correction signal that adjusts the microwave frequency. We find that digitizing the correction signal and using computer control of the microwave generator frequency yields good performance. The slope of the frequency modulated signal is used as a voltage-to-frequency conversion factor.

We also attempted to implement faster microwave frequency control using rf analog devices, but the results were not satisfactory. In this design, the frequency correction voltage was applied to the input of a voltage controlled oscillator (VCO) with a frequency span of 0.4 GHz to 1.3 GHz and a variable sensitivity as a function of applied voltage (from 63 MHz/V to 22 MHz/V). In-phase and quadrature signals were then mixed with a static-frequency microwave signal to produce a single-side-band at the NV$^-$ resonance frequency. We encountered several difficulties with this VCO-and-mixer design. Primarily, the VCO and mixer both have limited frequency ranges, which limits the measurable field range. Also, we experienced unreliable peak-locking, which we attributed to closely spaced, unwanted microwave harmonics at various microwave frequencies.

## Results and discussion:

First, we compare results obtained using the rate meter with results from the more conventional method of direct pulse counting. We measured amplitude modulated cw-ODMR spectra at a fixed applied magnetic field, and recorded data for the two methods simultaneously, using copies of the same pulse train. In this measurement, microwave power was modulated at 50 Hz: ON for 10 ms and OFF for the subsequent 10 ms in each cycle. The average emitted photon rate was 300 ms$^{-1}$ and average incident laser power was 112 μW. ODMR spectra from both measurements were normalized, as shown in Figure 2. The blue, square points are from the conventional photon counting method and the black, circular points are from the rate meter method. As shown in Figure 2, the signal-to-noise ratios (SNR) from the two methods are virtually identical.

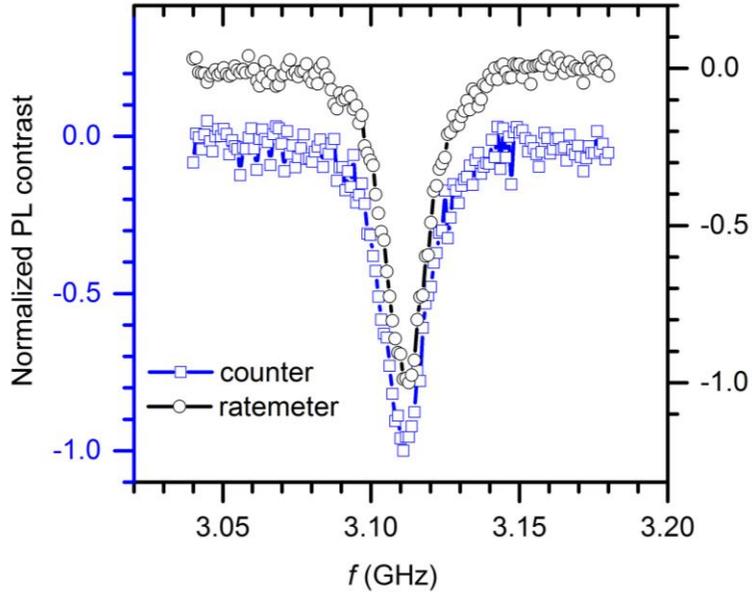

*Fig. 2 Amplitude-modulated, cw-ODMR spectra measured using both the differential rate meter and the traditional counter method on identical input signals. Data sets are offset for better visibility. Black circles represent the data recorded using the rate meter and blue squares represent data recorded using pulse counters. Signal-to-noise ratios are virtually identical.*

We now turn to using the field-dependence of the ODMR signal for magnetometry. The noise floor of a field measurement, $\delta B$, can be determined from the signal noise, $\delta S_{\text{FM}}$, of a frequency-modulated measurement using

$$\delta B = \frac{dB}{df}\left(\frac{dS_{\text{FM}}}{df_c}\right)^{-1}\delta S_{\text{FM}}.$$

The differential sensitivity, $\frac{dS_{\text{FM}}}{df_c}$, is obtained from the frequency modulated cw-ODMR spectrum as shown in Figure 3a. In this measurement, microwaves were frequency-modulated using a square wave. The modulation rate was 200 Hz and the amplitude was 5 MHz. Rate meter A received all the photons when the microwave frequency was high, ($f_1 = f_c + 5$ MHz) and rate meter B received all the photons when the microwave frequency was low ($f_2 = f_c - 5$ MHz). The normalized difference was measured as function of the center microwave frequency ($f_c$) for an integration time of 1 s per point. The noise level was determined from the off-resonance background, as shown in the dotted boxes. The histogram of the background signal is shown in Figure 3b; the standard deviation is $\delta S_{FM} = 0.65$ mV with a bandwidth of 1 Hz. We also determined the frequency sensitivity from a linear fit of the data points around the zero-crossing. The fit is shown as a red-line in Figure 3c. The measured slope $\frac{dS_{FM}}{df_c}$ for this fit is 3.96 x 10$^{-6}$ mV/Hz. The measured frequency sensitivity was 165 kHz/Hz$^{1/2}$, translates via the gyromagnetic ratio into a field sensitivity of 5.85 µT/Hz$^{1/2}$.

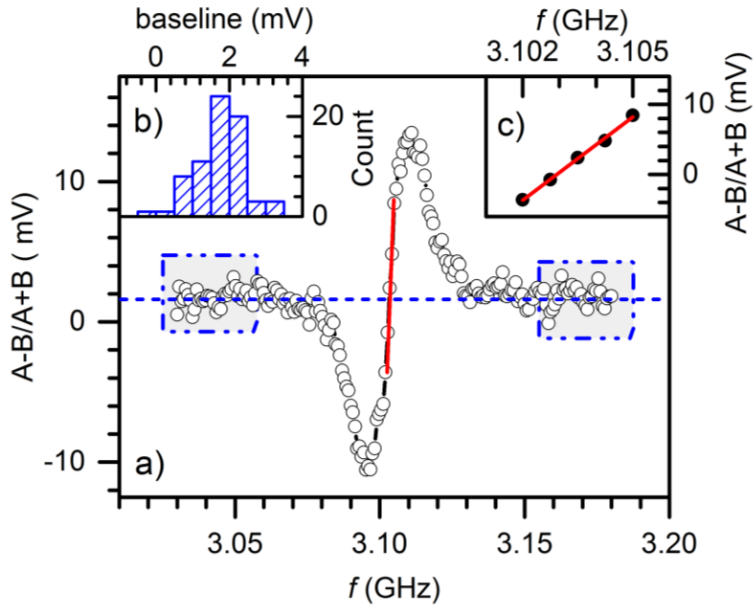

*Fig. 3 a) Frequency modulated cw-ODMR spectrum of NV$^-$ centers measured using the differential rate meter. The rate of frequency modulation was 200Hz with an amplitude of 5MHz. b) Histogram of the background noise taken from the data points shown as dotted box in fig.3a. c) A fit to the points nearest the zero crossing in a).*

Next, we demonstrate the peak tracking method for a wide-range magnetic field measurement. In this measurement, we lock and follow the magnetic resonance peaks of NV$^-$ centers from near zero magnetic field to 35 mT, revealing the Zeeman splitting of the NV$^-$ center as shown in Figure 4a. At magnetic fields near zero, the tracking feedback loop becomes unstable as the two resonance peaks overlap. The magnetic field (bottom axis of fig.4a) was independently measured using a Hall probe placed near the sample. The result in figure-4a is in agreement with the expected Zeeman splitting, and also shows the linearity of the probe and the detection scheme over wide a magnetic field range. The intercepts from the linear fits are virtually identical: 2.868 GHz and 2.869 GHz in agreement with the expected zero field splitting of the NV$^-$ center [31]. The uncertainty estimates from the standard deviations of the fit parameters are approximately 0.7 MHz, but systematic errors such as the field offset of our Hall probe may dominate. Figure 4b and 4c display the corresponding fit residual. Figure 4d is the portion of the upper plot (shown in dotted box) from 20 mT to 22 mT, showing individual data points.

In principle, the potential field range of this measurement scheme is determined by the frequency range of microwave components such as the microwave frequency source, RF cables and the microstrip antenna. In this demonstration, however, we are limited by the maximum field of our small, air-cooled electromagnet.

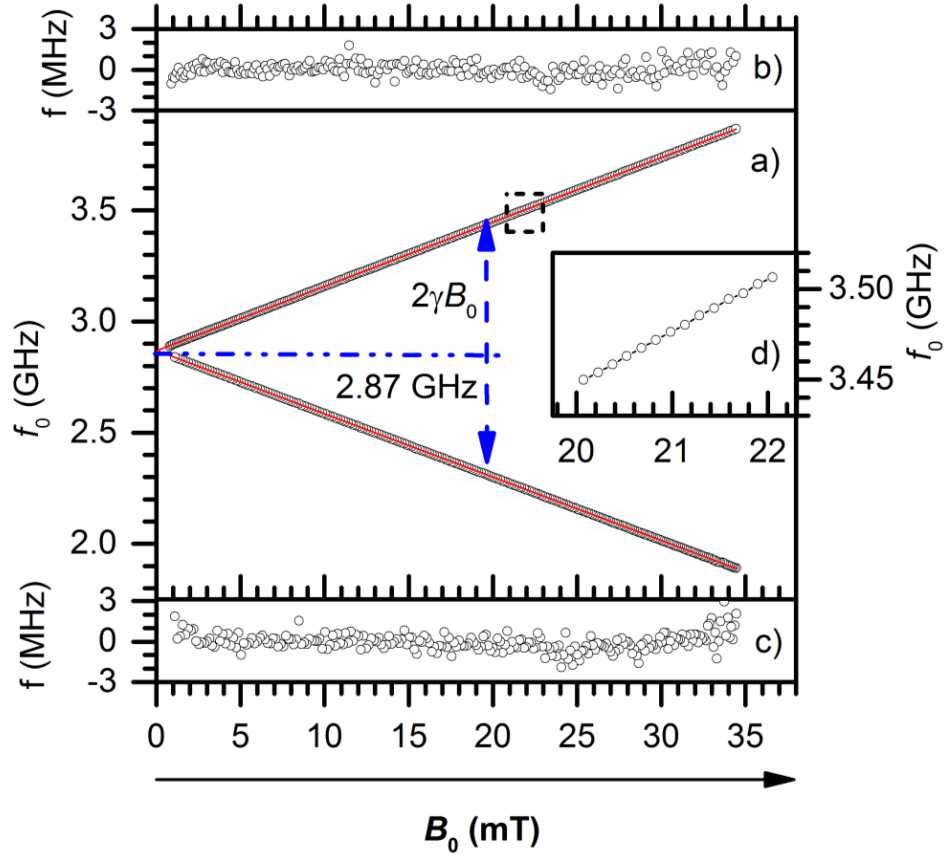

*Fig4. Continuous magnetic field measurement using custom-built rate meter and PID control for peak tracking. Figure-4a shows the Zeeman effect of NV- centers, measured by tracking the resonance peaks. Figures 4b and 4c plot the residual of linear fits. Figure 4d is a magnified portion of the upper curve outlined by the dotted box from 20 mT to 22 mT.*

To demonstrate real-time data processing, frequency control, and field tracking of the feedback scheme, we applied an AC magnetic field with frequency 0.1 Hz and amplitude 300 µT. In Fig. 5, the square, blue data points are the applied fields measured using a Hall probe. The round, black data points are the resonance frequencies determined by peak tracking. As shown in figure 5, the feedback loop locks and follows the AC magnetic field in real-time, demonstrating reliable field tracking at a sweep rate of ≈ 110 µT/s. Tracking was also achieved at higher rates up to 220 µT/s, but not reliably.

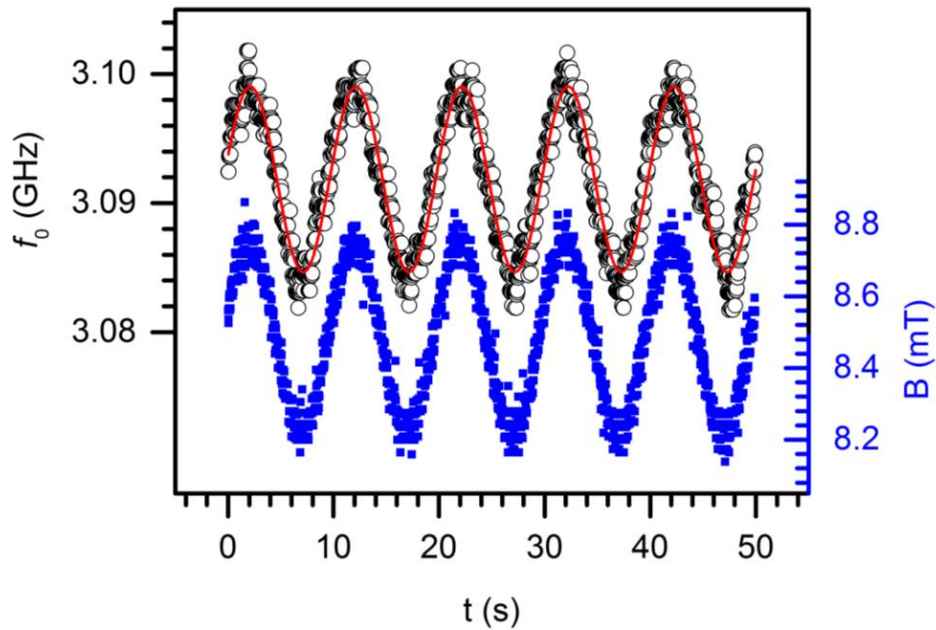

*Figure 5. Real time data processing and frequency control using a rate meter and feedback control. The square, blue points are magnetic field measured using a Hall probe. The round, black points are the tracked frequencies of the feedback loop. The red curve is a sinusoidal fit of the tracked frequency changes. The feedback loop tracks a 0.1 Hz AC magnetic field of amplitude 300 µT. The maximum rate of magnetic field sweep is ≈ 110 µT/s.*

## Conclusion:

In conclusion, we implemented a frequency modulated differential photon rate measurement for optically detected magnetic resonance of NV- centers using single photon detectors. We demonstrated that the rate meter method provides virtually identical signal-to-noise ratios as compared to conventional photon-counting methods. Using differential measurement and active feedback (PID) control, we lock and follow the magnetic resonance peak of NV- centers from near-zero fields to 35 mT. The sensitivity and maximum tracked-field sweep rate of our detection scheme are 6 µT/√Hz and 110 µT/s respectively. The implementation of our peak-tracking method is relatively simple and could easily be integrated with other experimental setups such as scanning probe microscopy for nanoscale magnetometry. Also, the described rate meter can detect photon rates from 2 ms$^{-1}$ to 500 ms$^{-1}$. Therefore, this method is well suited for use in single NV$^{-}$ center as well as small ensembles of NV$^{-}$ centers.

## Acknowledgements:

K. Ambal acknowledges support under the Cooperative Research Agreement between the University of Maryland and the National Institute of Standards and Technology Center for Nanoscale Science and Technology, Award 70NANB14H209, through the University of Maryland. The authors especially thank Alan Band and David M. Rutter for design and implementation of the ratemeter instrumentation and for their mentoring in successful completion of the project.

## Contributions:

K.A. designed and performed the experiment, measurement analysis and contributed to the manuscript preparation. R.D.M and K.A. conceived the concept. R.D.M oversaw the project including analysis, discussion and conceptual guidance and he contributed to the manuscript preparation.

## Competing interests:

Based on the real-time magnetic resonance peak tracking concept using ratemeter discussed in this study, an invention disclosure was made at the technology commercialization office of the University of Maryland.